\documentclass[%
reprint,
superscriptaddress,
amsmath,
amssymb,
aps,
prb,
]{revtex4-2}

\usepackage[final]{changes}
\usepackage{graphicx}
\usepackage{dcolumn}
\usepackage{bm}

\begin{document}

\title[Magnetic skyrmion interactions in the micromagnetic framework]{Magnetic skyrmion interactions in the micromagnetic framework}

\author{R. Brearton}
 \email{richard.brearton@physics.ox.ac.uk}
 \affiliation{Department of Physics, Clarendon Laboratory, University of Oxford, Oxford, OX1 3PU\\}
 \affiliation{Magnetic Spectroscopy Group, Diamond Light Source, Fermi Ave, Didcot OX11 0DE, England}

\author{G. van der Laan}
 \email{gerrit.vanderlaan@diamond.ac.uk}
 \affiliation{Magnetic Spectroscopy Group, Diamond Light Source, Fermi Ave, Didcot OX11 0DE, England}

\author{T. Hesjedal}
 \email{thorsten.hesjedal@physics.ox.ac.uk}
 \affiliation{Department of Physics, Clarendon Laboratory, University of Oxford, Oxford, OX1 3PU\\}

\date{\today}

\begin{abstract}

Magnetic skyrmions are localized swirls of magnetization with a non-trivial topological winding number. This winding increases their robustness to superparamagnetism and gives rise to a myriad of novel dynamical properties, making them attractive as next-generation information carriers. Recently the equation of motion for a skyrmion was derived using the approach pioneered by Thiele, allowing for macroscopic skyrmion systems to be modeled efficiently. This powerful technique suffers from the prerequisite that one must have a priori knowledge of the functional form of the interaction between a skyrmion and all other magnetic structures in its environment. Here we attempt to alleviate this problem by providing a simple analytic expression which can generate arbitrary repulsive interaction potentials from the micromagnetic Hamiltonian. We also discuss a toy model of the radial profile of a skyrmion which is accurate for a wide range of material parameters.

\end{abstract}

\maketitle

\section{Introduction}

Magnetic skyrmions (skyrmions hereafter) have recently been the subject of great interest due both to their emerging physical effects and potential for applications \cite{nagaosa2013topological, everschor2018perspective}. Skyrmionic devices promise exceptional electrical efficiency -- skyrmions can be directly driven by ultra-low current densities, and even controlled in the absence of current by static magnetic field gradients \cite{yu2012skyrmion, zhang2018manipulation}. Their small size and unusual dynamics allow for their utilization in conventional computers, which has prompted a wide study of skyrmionic re-implementations of arithmetic logic units and memory storage devices \cite{luo2018reconfigurable, fert2013skyrmions}. Additionally, recent studies suggest that skyrmions could find use in exotic devices such as stochastic and reservoir computers \cite{pinna2018skyrmion, prychynenko2018magnetic}.

In order to study such devices, one commonly employs the theory of computational micromagnetism. This involves the numerical integration of the Landau-Lifshitz-Gilbert (LLG) equation, typically using the popular finite differences method \cite{vansteenkiste2014design, donahue1999oommf}. In this technique, the fundamental unit of information operated on numerically is a volume of magnetization. As a very large number of these magnetization volumes is required to construct a skyrmion, this technique can quickly become computationally infeasible when systems containing a large number of skyrmions are of interest. 

Another approach to studying skyrmion dynamics involves integrating Thiele's equation, which has recently been extended to describe skyrmion dynamics \cite{thiele1973steady, lin2013particle}. Thiele's equation can be written for the $i^\mathrm{th}$ skyrmion in a system as
\begin{equation}
\vec{G}\times\dot{\vec{x}}_i - D\alpha\dot{\vec{x}}_i = 
-\nabla E_i(\vec{x}_i),
\label{eq:Thiele}
\end{equation}
\noindent
where $\vec{G}$ is the gyromagnetic coupling vector, $D$ is the dissipative tensor, $\alpha$ is the damping coefficient, and 
$E_i = U_{i,1} + U_{i,2} + \cdots$ 
is the sum of all of the interaction potentials affecting the $i^\mathrm{th}$ skyrmion. As $\vec{x}_i$ represents the position of a skyrmion, numerical integration of Eq.~(\ref{eq:Thiele}) directly solves for the path traversed by a skyrmion, completely avoiding any consideration of the skyrmion's constituent magnetic moments. This significant simplification allows simulations to be run which study length scales which were previously inaccessible, provided that all interaction potentials appearing in $E_i$ are understood.

While the functional form of the skyrmion-skyrmion interaction potential has been known for some time \cite{lin2013particle}, calculating the interaction between a skyrmion and other spin textures which co-exist with the skyrmion state has remained a difficult task. In particular, it is widely advertised that skyrmions repel from the edge of the materials in which they exist --- this so-called edge repulsion motivated the design of skyrmion racetrack memory and is critical to most device schematics \cite{zhang2015skyrmion}. In spite of this, various groups have derived drastically different functional forms for the skyrmion-edge repulsion interaction potential. The seminal work by Meynell \emph{et al.}~\cite{meynell2014surface}, in which so-called `surface twist instabilities' are characterized in detail theoretically, predicts that the skyrmion-surface twist interaction potential should be proportional to an exponential. Here it is found that the argument of the exponential provided leads to a poor fit to micromagnetic calculations. In the stochastic computing study by Pinna \emph{et al.}~\cite{pinna2018skyrmion} an accurate interaction potential is derived by over-constraining the potential and fitting a product of six exponentials to micromagnetic data, treating the argument of each exponential as an independent fitting parameter. 

\begin{figure}[h]
\includegraphics[width=8.636cm]{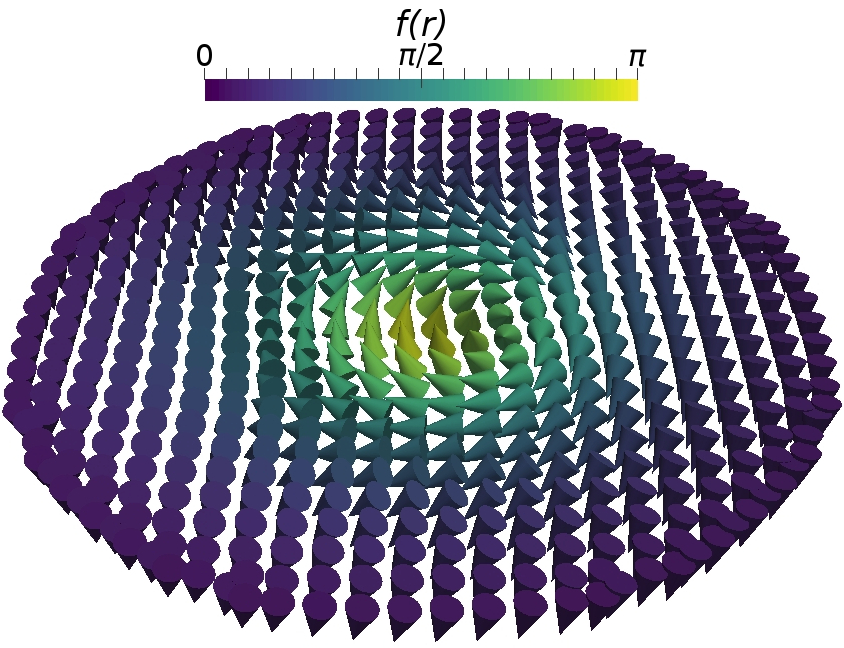}
\caption{\label{fig:f_r_skyrmion} 
Visualization of an isolated skyrmion, where the direction of magnetization is represented by cones. These cones are colored by the angle between the magnetization and the direction of the out-of-plane magnetic field --- this angle is referred to as $f(r)$ throughout the text.
}
\end{figure}

In this paper we derive a general form of the interaction potential between any two repulsive, localized micromagnetic objects. The special cases of the skyrmion-skyrmion and skyrmion-edge interactions are discussed, and a correction to the previously derived skyrmion-edge interaction potential is provided.
In order to study the rich physics present in skyrmion hosting systems, it is often necessary to analytically construct a skyrmion. The task of writing down the angle between an isolated skyrmion's magnetization vector and the out-of-plane magnetic field is seemingly simple, but drastically complicated by the nonlinear differential equation that constrains it. This angle, hereafter referred to as $f(r)$, is visualized in Fig.~\ref{fig:f_r_skyrmion}. Here the previous approaches to writing down $f(r)$ analytically are discussed and a simple, accurate toy model of a skyrmion's radial profile is provided. 

\section{Theory}
\subsection{Skyrmions from the energy functional}
In the absence of the dipolar interaction, the micromagnetic energy functional is given by \cite{thiaville2012dynamics, cortes2018proposal}
\begin{equation}
E = \int_{V} \left[
J\left(\nabla \vec{m} \right)^2 + 
D \vec{m} \cdot \left( \nabla \times \vec{m} \right) - 
\vec{m} \cdot \vec{B} 
\right]d^3 \vec{r} \;,
\label{eq:micromag}
\end{equation}
\noindent
where the local magnetization is given by $\vec{m}$.  $\vec{B}$ is the external magnetic flux density,  $J$ and $D$ are the exchange and Dzyaloshinskii-Moriya constants, respectively. Any experimentally observable magnetic configuration should be a local minimum of Eq.~(\ref{eq:micromag}). The Euler-Lagrange (EL) equations of Eq.~(\ref{eq:micromag}) have been studied extensively (see e.g.\ the work of Melcher \cite{melcher2014chiral}) and it has been shown that, for a system with a $B$-field oriented along the $z$-axis, an isolated skyrmion magnetization configuration defined by the ansatz
\begin{equation}
\frac{\vec{m}_{\mathrm{Sk}}\left( \vec{r} \right)}{m_\mathrm{S}} = \left( C \sin{\left[f(r)\right]}, S \sin{\left[f(r)\right]}, \cos{\left[f(r)\right]} \right)\;,
\label{eq:m(x)}
\end{equation}
\noindent
is a solution for $C = \cos\left(n\theta + \gamma\right)$ and $S = \sin\left(n\theta + \gamma\right)$ with $m_\mathrm{S}$ the saturation magnetization and $\theta$ the polar angle. The helicity $\gamma$ determines the type of the skyrmion, where $\gamma=0, \pi$ corresponds to N\'{e}el-type solutions and $\gamma=\pm \pi/2$ yields Bloch-type solutions \cite{nagaosa2013topological}. In order to minimize the micromagnetic Hamiltonian given in Eq.~(\ref{eq:micromag}) one must set the skyrmion winding number to $n$ = 1 \cite{foster2018composite}. Additionally, the radial EL equation of Eq.~(\ref{eq:micromag}) constrains $f(r)$ to satisfy \cite{lin2013particle}
\begin{eqnarray}
Jrf'' + Jf' + D\sin \gamma \left(1-\cos{2f}\right)   \nonumber\\
- \frac{J}{2r}\sin{2f} - Br\sin{f} = 0 \;,
\label{eq:f(r)}
\end{eqnarray}
\noindent
where $f'=\partial f/\partial r$.

One can solve this equation numerically for $f(r)$, but some important information can be extracted analytically. Firstly, Eq.~(\ref{eq:f(r)}) contains the solution $f=0$: the field polarized state is a solution to the EL equations. Secondly, far from the core of the skyrmion we expect to retrieve the field polarized state, so that $\lim_{r\to\infty}\left(f\right) = 0$. In this limit $\left(1-\cos2f\right) \approx 0$ reducing Eq.~(\ref{eq:f(r)}) to the modified Bessel equation, with solutions $K_i(r) \propto e^{-r/\lambda}/\sqrt{r}$ for $\lambda = \sqrt{J/Bm_\mathrm{S}}$. As the skyrmion magnetization tends towards the field-polarized state exponentially quickly, the skyrmion lattice state can be retrieved by noticing that linear superpositions of well-separated skyrmion configurations asymptotically approach solutions of the EL equations.

\subsection{Energy out of equilibrium}

In order to calculate interaction potentials, it is lucrative to consider the effect of superposing two arbitrary localized magnetization configurations, $m_1$ and $m_2$, on the energy of the system. As argued above, if the two magnetization textures are localized over a length scale $\Lambda$, in the limit $R \gg \Lambda$ the superposition of the two magnetization textures $\vec{m}_{\mathrm{total}} = \vec{m}_1( \vec{r} ) + \vec{m}_2( \vec{r} + \vec{R} )$ approaches a minimum of the micromagnetic energy functional. As $R$ becomes comparable in size to $\Lambda$ then $\vec{m}_1$ and $\vec{m}_2$ will begin to interfere with each other --- \added{drawing an analogy to classical mechanics by} writing 
$E = \int_V \mathcal{E}\left(\vec{m}(\vec{r}), \vec{m}'(\vec{r})\right)d^3\vec{r}$
the energy of the system will increase by an amount given by the calculus of variations as \cite{landau2013mechanics}
\begin{equation}
\Delta E =
\int_V \delta \vec{m}_\mathrm{total} \cdot \left[ 
\frac{\partial\mathcal{E}}{\partial \vec{m}} - 
\frac{d}{d\vec{r}}\left( \frac{\partial\mathcal{E}}{\partial \vec{m}'} \right)
\right] d^3\vec{r}\;,
\end{equation}
\added{where primes denote spatial derivatives.} This expression is difficult to work with because of the presence of the unknown $\delta\vec{m}_\mathrm{total}$ term. It is critical to understand that this increase in energy arises due to the non-negligible overlap between $\vec{m}_1$ and $\vec{m}_2$ for small $R$. In order to calculate this overlap we define the difference between the $\vec{m}_i$ and the field-polarized `background' as $\vec{\mu}_i =  \vec{m}_i - \left(0,0,m_\mathrm{S}\right)$. \added{One can then measure the overlap between two adjacent magnetization configurations using} \deleted{The overlap between $\vec{m}_1$ and $\vec{m}_2$ is then simply given by}
\added{
\begin{equation}
\psi = \int_V \left| \vec{\mu}_1(\vec{r}) \times \vec{\mu}_2(\vec{r} - \vec{R}) \right| d^3 \vec{r}\;.
\label{eq:overlap}
\end{equation}
}
\added{To understand this construction, it is useful to consider the limiting behaviour of the integrand. An important property of the $\vec{\mu}_i(\vec{r})$ is that they are localized; $\lim_{|r|\to\infty}\left|\vec{\mu}_i(\vec{r})\right| = 0$ implies that $|\vec{R}|$, the separation distance between the two textures, must be small for the integrand to be non-negligible anywhere. Additionally the cross product weights the integrand by the angle between the magnetization textures -- as $\vec{m}_1$ and $\vec{m}_2$ diverge, the angle between $\vec{\mu}_1$ and $\vec{\mu}_2$ increases (to a maximum of 90$^\circ$). As such, e}\deleted{E}quation~(\ref{eq:overlap}) provides a convenient parameter with which one can describe the energy increase in the system\added{ due to magnetization overlap}. In particular, for\added{ very} large values of $\psi$ the internal structure of $\vec{m}_1$ will be affected by the tail of $\vec{m}_2$ and vice versa --- this is precisely the limit in which the Thiele equation is no longer valid \cite{thiele1973steady}. The Taylor expansion of $\Delta E (\psi)$ then reads
\begin{equation}
\Delta E = K \psi + \mathcal{O}\left( \psi^2 \right)\;,
\label{eq:inc_energy}
\end{equation}
\noindent
where $K$ is a constant of proportionality which can be treated as a fitting parameter and determined using micromagnetic simulations. As $R$ decreases, $\vec{m}_1$ and $\vec{m}_2$ will in general deform nonlinearly, but in the large $R$ limit, Eq.~(\ref{eq:inc_energy}) provides an accurate measure of the interaction potential between $\vec{m}_1$ and $\vec{m}_2$. It is worth emphasizing that in this section the discussed magnetic textures have been kept arbitrary and that this approach should determine the strength of the interaction between arbitrary\deleted{,} well-separated localized magnetic textures.

\section{Results}

\subsection{Toy model of a skyrmion}

To evaluate Eq.~(\ref{eq:overlap}) with some accuracy, it is necessary to construct an analytical approximation of the solution to Eq.~(\ref{eq:f(r)}) which is asymptotically exact and finite for all $r$. Previous parameterizations of this $f(r)$ typically concern themselves with accuracy within a certain limit --- varying linearly for small $r$ or decaying like a modified Bessel function for large $r$. When considering the behavior of the skyrmion core, linear or piece-wise linear models have been used in the past \cite{kiselev2011chiral, jiang2017direct}. To study the far-field behavior often the linear component is dropped entirely and a pure modified Bessel function is used \cite{lin2013particle, foster2018composite}, while in some cases trigonometric approximations which capture some of the near and far-field behavior are employed \cite{romming2015field, leonov2016properties, wang2018theory}. 

It is possible to construct an accurate model of $f(r)$ by smoothly continuing the short-ranged solution to the asymptotic solution. This can be done using a modified logistic function $L(r) = A/[1 + e^{-\kappa r(1-r_0)/(J/D)}] - A/[[1 + e^{-\kappa r(1+r_0)/(J/D)}] $ which has been adjusted to meet the requirement $L(0)=0$. This yields the much more accurate expression
\begin{equation}
f(r) = [1-L(r)](\pi - kr) + L(r)\frac{e^{-r/\lambda}}{\sqrt{r}}.
\label{eq:toy_model}
\end{equation}

The purpose of $A$, $\kappa$, $r_0$, and $k$ can be understood intuitively. $A$ is the coefficient of the modified Bessel function of the second kind, which is the particular solution to Eq.~(\ref{eq:f(r)}) for large $r$ which is relevant for skyrmions. The parameter $\kappa$ parameterizes the smoothness with which $f(r)$ varies between short and long range solutions -- $\lim_{\kappa\to\infty}L(r)=H(r-r_0)$ where $H(r-r_0)$ denotes the Heaviside step function. This statement elucidates the role of $r_0$, which denotes the radius from the skyrmion beyond which the radial profile is better described by a modified Bessel function than a linear function. The gradient of the linear part of the profile near the origin is given by $k$.

\begin{figure}[b]
\includegraphics[width=8.636cm]{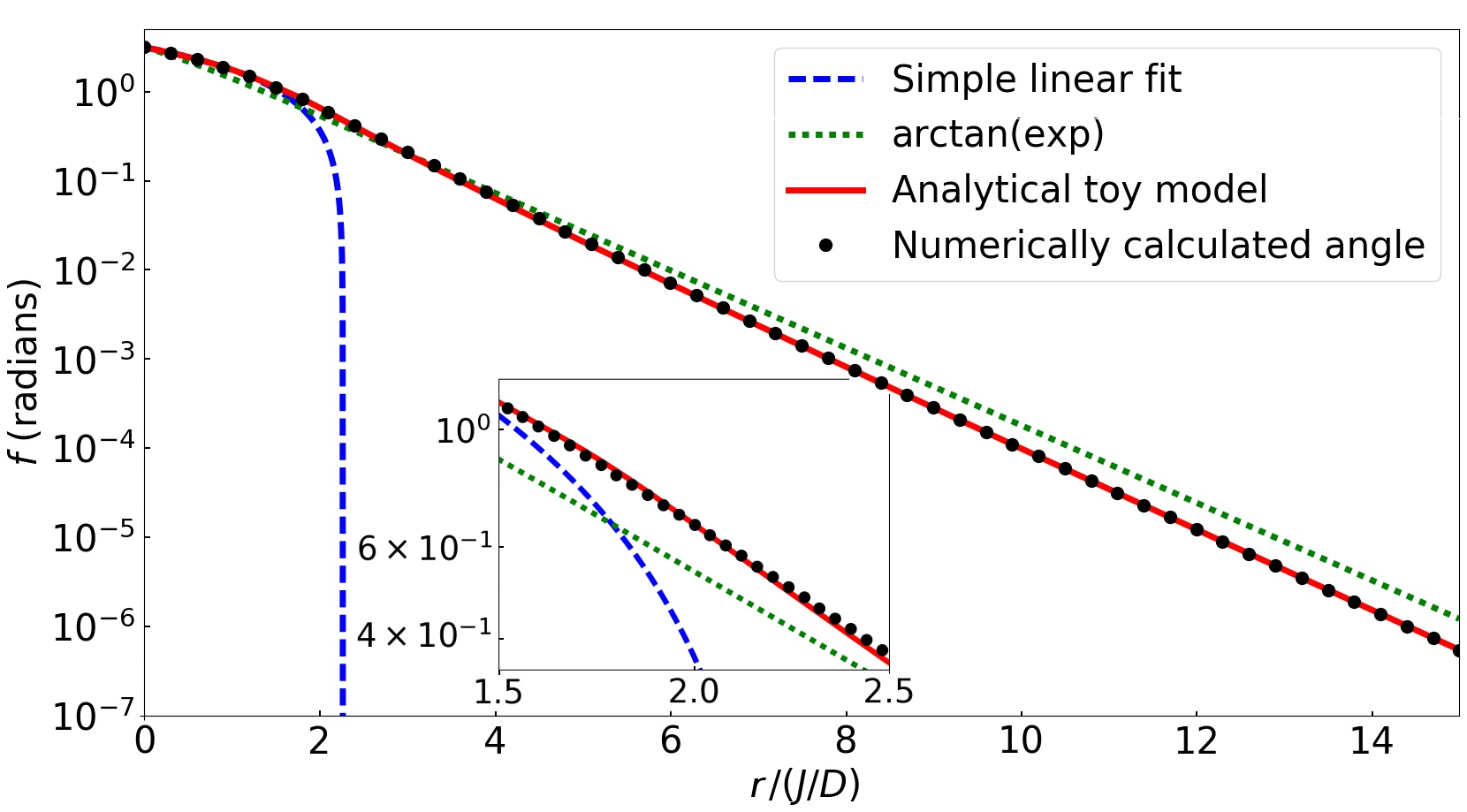}
\caption{\label{fig:toy_model} 
A fit of the analytical toy model described in Eq.~(\ref{eq:toy_model}) to the numerical solution of Eq.~(\ref{eq:f(r)}). In the differential equation the values of $J$, $D$, and $B$ were taken to be 1 while choosing $\gamma=-\pi/2$. The toy-model parameters were refined to be $A=6.95$, $\kappa=5.54$, $r_0=1.61$, $k=1.37$, and $\lambda=1$. For comparison, the popular piecewise linear model is plotted \cite{kiselev2011chiral, jiang2017direct}, as well as the model $f(r) = 4\arctan (e^{-\frac{r}{\lambda}})$ \cite{leonov2016properties}.
}
\end{figure}

A fit of this model to the numerical solution of Eq.~(\ref{eq:f(r)}) is given in Fig.\ \ref{fig:toy_model}, alongside a comparison with previous work. As we simply require that any model of $f(r)$ used to derive an interaction potential is asymptotically exact and finite near the origin, none of the parameters introduced in this model explicitly appear in expressions relating to interaction potentials. In principle they could be calculated, but for the remainder of this work they are kept arbitrary.

\subsection{Skyrmion-skyrmion interactions}

In order to derive a simple analytical form for the skyrmion-skyrmion interaction potential $U_{\mathrm{SkSk}}$, one can substitute the toy model for $f(r)$ into the skyrmion magnetization configuration given in Eq.~(\ref{eq:m(x)}), which can then be substituted into Eq.~(\ref{eq:overlap}). The product \added{$\left|\vec\mu_{\mathrm{Sk}}(\vec{r}) \times \vec\mu_{\mathrm{Sk}}(\vec{r}-\vec{R})\right|$} has a maximum at $\vec{r}=\vec{R}/2$ and symmetrically decreases about this point. The symmetry of the skyrmion along its axis reduces the the volume integral in Eq.~(\ref{eq:overlap}) to an area integral which can be directly calculated numerically. To carry out this integral analytically it is necessary to make a series of simplifying approximations. One can consider the area integral \added{$\int_A \left|\vec{\mu}_1(\vec{r}) \times \vec{\mu}_2(\vec{r} - \vec{R})\right| d^2 \vec{r}$} as an infinite sum of one dimensional line integrals which connect the two skyrmions, whose largest contribution comes from the path that directly connects the two skyrmions. In a new set of coordinates $\beta = r - R/2$, choosing this path to be along the azimuthal angle $\theta=0$, Taylor expanding and keeping only highest order terms in the integrand, this one dimensional integral is given by
%
%
\begin{eqnarray}
\hspace*{-0.5cm}
U_{\mathrm{SkSk}}(R) \propto e^{-\frac{R}{\lambda}}\int^{\frac{R}{2}}_{-\frac{R}{2}}{ d\beta \frac{L(\beta + \frac{R}{2}) [1 - L(\beta-\frac{R}{2})]} {\sqrt{\frac{R^2}{4} - \beta^2}}}.
\label{eq:sksk_integral}
\end{eqnarray}

While one can Taylor expand this integrand and integrate, the behavior of the potential for large $R$ is dominated by the rapidly decaying exponential term, re-deriving the well known result
\begin{equation}
U_{\mathrm{SkSk}}(R)\propto e^{-\frac{R}{\lambda}}.
\label{eq:SkSk_interaction}
\end{equation}

\subsection{Skyrmion-boundary interactions}

The calculation of the interaction between a surface twist and a skyrmion follows in much the same way as above. The magnetization profile of a surface twist in a semi-infinite film bounded by the $x$-axis is given by
\begin{equation}
\frac{\vec{m}_{\mathrm{Twist}}}{m_\mathrm{S}} = \left( 0, \sin{\theta(x)}, \cos{\theta(x)}  \right),
\label{eq:surface_twist}
\end{equation}
\noindent
where $\theta(x)$ denotes the tilt angle of the spins with respect to the direction of the out-of-plane magnetic field, which has been taken to be along the $z$-axis. The full form of $\theta(x)$ has been derived analytically, with asymptotic behavior described by  $\theta(x)\propto e^{-x/\lambda}$ \cite{meynell2014surface}. As such the integral
\added{$\int_{A} \left|\vec\mu_{\mathrm{Twist}}(\vec{r}) \times  \vec\mu_{\mathrm{sk}}(\vec{r}-\vec{x})\right|d^2\vec{r}$} yields the same result to lowest order
\begin{equation}
U_\mathrm{SkTwist} \propto e^{-\frac{x}{\lambda}},
\label{eq:SkTwistPotential}
\end{equation}
\noindent
where now the magnitude of the interaction potential depends on the distance between the skyrmion and the boundary of the material, which is in this geometry given by the $x$-coordinate of the skyrmion. As such the interaction between a skyrmion and the boundary of a material can be considered to be equivalent to the interaction between the real skyrmion and virtual skyrmions centered on the boundary. In this sense one can understand how skyrmions interact with different shaped boundaries --- in a convex boundary more of the virtual skyrmions are further away from the real skyrmion (reducing the magnitude of the prefactor in Eq.~(\ref{eq:SkTwistPotential})) while the opposite is true near to a concave boundary.

\subsection{Micromagnetic calculations}

For the purposes of fitting and benchmarking the results of the previous sections, it is necessary to carry out micromagnetic calculations of the interaction potentials of interest. Typically this is done by initializing two skyrmions near to each-other and tracking the motion of each skyrmion to calculate the force, which is integrated to derive the potential \cite{pinna2018skyrmion, foster2018composite}. The out of equilibrium initialization of the system leads to the technical issue that only the final state of the simulation, in which the skyrmion is at rest, represents a physical magnetization texture which minimizes Eq.~(\ref{eq:micromag}). 

\begin{figure}
\includegraphics[width=8.6cm]{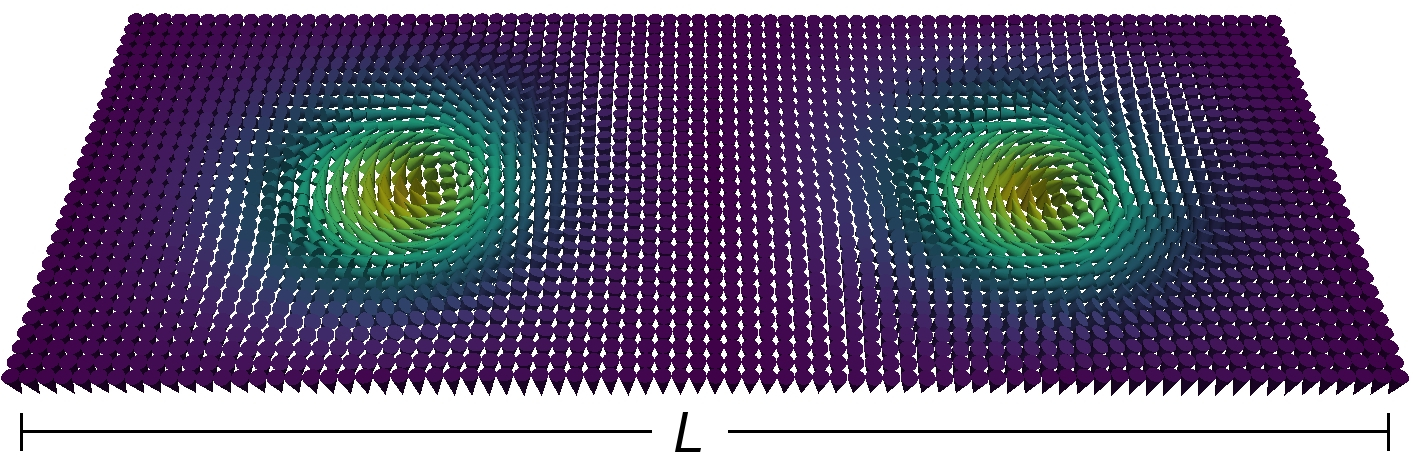}
\caption{ 
An example of a micromagnetic calculation performed to determine the skyrmion-skyrmion interaction potential. The length $L$ was progressively shortened to map out the energy increase of the system due to interactions between the skyrmions.
}
\label{fig:f_r_two_skyrmions}
\end{figure}

\begin{figure}
\includegraphics[width=8.636cm]{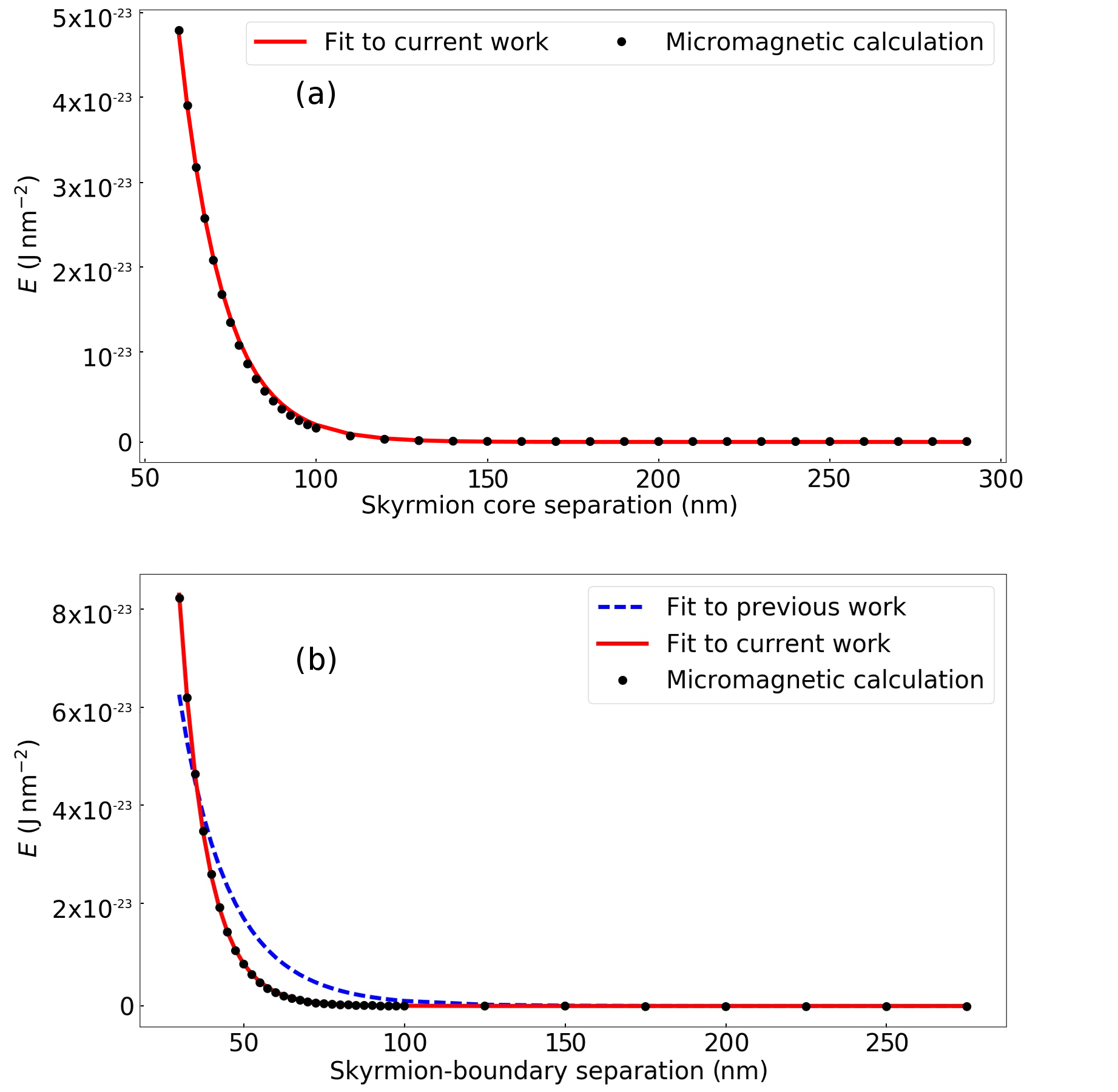}
\caption{ 
Fit of results derived in this work to micromagnetic calculations of interaction potentials. 
(a) Fit of Eq.~(\ref{eq:SkSk_interaction}) to the micromagnetically calculated skyrmion-skyrmion interaction potential. 
(b) As in (a), but including a fit of the skyrmion-surface twist interaction potential as derived in \cite{meynell2014surface} for comparison.
}
\label{fig:Micromagnetic_fits}
\end{figure}

In this work we propose a simple and more accurate solution, in which two skyrmions are confined in a film of length $L$ which is made progressively shorter and shorter -- the simulation geometry is shown in Fig.~\ref{fig:f_r_two_skyrmions}. The total energy of the relaxed system is given by $E = E_{\mathrm{FP}} + E_{\mathrm{Sk}} + U_{\mathrm{SkSk}}$, where $E_{\mathrm{FP}}$ is the energy of the field-polarized state, $E_{\mathrm{Sk}}$ is the internal energy of a skyrmion, and $U_{\mathrm{SkSk}}$ is the skyrmion-skyrmion interaction potential. The energies $E_\mathrm{Sk}$ and $E_\mathrm{FP}$ can be calculated in advance for the material parameters of interest. As $L$ is decreased and the two skyrmions are forced closer to each other, the increase in $U_{\mathrm{SkSk}}$ describes the interaction potential. The calculated $U_{\mathrm{SkSk}}$ is given in Fig.~\ref{fig:Micromagnetic_fits}(a) and compared to Eq.~(\ref{eq:SkSk_interaction}).

In order to determine the interaction potential between a skyrmion and a surface twist, a similar technique was utilized in which a skyrmion was relaxed in a film which was periodic along the $y$-direction but finite sized along the $x$-direction. The equilibrium energy was calculated for increasingly small $x$ and is shown in Fig.\ \ref{fig:Micromagnetic_fits}(b), compared both to Eq.~(\ref{eq:SkTwistPotential}) and the functional form of the interaction potential provided in Ref.\ \cite{meynell2014surface}.

\section{Discussion}

We have derived a functional which can generate interaction potentials between arbitrary localized magnetic states which, when superposed, repel and do not form a bound state. This expression has been used to derive the skyrmion-skyrmion interaction potential, which was found to be consistent with previous work. We have demonstrated the predictive power of this model by providing a significant correction to the skyrmion-surface twist interaction potential and explained the result in terms of interactions with virtual skyrmions -- this has been verified by means of micromagnetic simulations. \added{While the discussions present in this article focus on two-particle interactions, the linearity of the Thiele equation allows many-particle interactions to be accounted for by linearly superposing the force due to each additional particle.} It bears noting that the above discussion has taken place in the limited context of micromagnetism --- while the Thiele equation provides useful limits over which the interaction potential needs to be valid, it is likely that such an approach would hold in other nonlinear field theories. 

In order for the calculated interaction potentials to be accurate, a toy model of the radial profile of a skyrmion has been provided. While only simple functional forms for interaction potentials have been derived in this work, using the provided skyrmion magnetization profile one could in principal derive higher order corrections to the expressions given here. We hope that this toy model will find use in a wide range of skyrmionic problems.

An important topic that has not been referred to up to this point is the phenomenon of skyrmion attraction. Skyrmions have been observed theoretically and experimentally to be attractive only in single crystalline samples which have thicknesses greater than or comparable to the helical wavelength \cite{leonov2016three, loudon2018direct}. In such systems the skyrmion state coexists with the conical state for much or all of its phase pocket and is modulated along its axis. In many ultra-thin films suitable for device applications there is no observed conical twisting, particularly in N\'{e}el-type systems where skyrmions often coexist with the field-polarized state in small out-of-plane magnetic fields. It is in this limit that the results presented in this work are most relevant. 

\added{
It is in principle possible to go beyond this limit and use the framework presented here to calculate interaction potentials between magnetization configurations embedded in, for instance, a conical background. The only required change would be that the form of the localized magnetization $\vec{\mu}(\vec{r})$ used in Eq.~(\ref{eq:overlap}) must become $\vec{\mu}_i (\vec{r}) = \vec{m}_i(\vec{r}) - \vec{m}_\mathrm{Conical}(\vec{r})$. 
}

\section{Conclusions}

A general form of the interaction potential between two repulsive localized micromagnetic textures has been derived, allowing for the use of the Thiele equation in a wide range of contexts. In order to work with this expression analytically, an accurate toy model of the radial profile of a skyrmion has been provided -- this toy model was compared with the results of previous work. The special cases of the skyrmion-skyrmion and skyrmion-surface twist interaction potentials were studied analytically and their accuracy was verified numerically. The derived skyrmion-surface twist interaction potential was found to be a significant correction to the previously derived expression. To assist with future energetic calculations, the numerical methods used to determine the interaction potentials have been discussed in detail.

\section*{Acknowledgments}

The authors acknowledge the work of Alexander Koziell-Pipe, whose micromagnetic investigations highlighted the need for a consistent mathematical treatment of skyrmion interactions. Financial support by the EPSRC (EP/N032128/1) is gratefully acknowledged.


%

\end{document}